\documentclass[epjCONF,columns]{svjour} 
\usepackage{wrapfig}
\usepackage[pdftex]{graphicx}
\usepackage{sidecap}
\usepackage{subfig}
\usepackage{graphics}
\usepackage[varg]{txfonts} % Times fonts
\usepackage[latin1]{inputenc}

\usepackage{natbib}
\usepackage{ref_macros}

\usepackage[english]{babel}
\usepackage{graphicx}
\usepackage{verbatim}
\usepackage{multirow}

\begin{document}
\title{Tidal Disruption Flares as the Source of Ultra-high Energy Cosmic Rays}
\author{ Glennys R. Farrar  ~~{\email{gf25@nyu.edu}} }
\institute{
Center for Cosmology and Particle Physics, Department of Physics, New York University, NY 10003, USA}
\abstract{
The optical spectral energy distributions of two tidal disruption flares identified by van Velzen et al. (2011) in archival SDSS data, are found to be well-fit by a thin-accretion-disk model.  Furthermore, the inferred Supermassive Black Hole mass values agree well with the SMBH masses estimated from the host galaxy properties.  Integrating the model SEDs to include shorter wavelength contributions provides an estimate of the bolometric luminosities of the accretion disks.  The resultant bolometric luminosities are well in excess of the minimum required for accelerating UHECR protons.  In combination with the recent observational estimate of the TDF rate (van Velzen and Farrar, these Proceedings), the results presented here strengthen the case that transient jets formed in tidal disruption events may be responsible for accelerating all or most UHECRs. 
} 

\maketitle

\section{Introduction}
It was proposed in ref. \cite{fg09} (FG09 below) that all or most UHECRs may be accelerated in transient jets produced by stellar tidal disruption events or extremely powerful disk instabilities in AGNs.  This proposal was motivated in part by the shortcomings of GRBs and powerful AGNs, the primary contenders for UHECR acceleration, and in part by the association reported by the Auger collaboration \cite{augerScience07} between UHECRs and galaxies in the Veron-Cetty Veron (VCV) list of possible AGNs\cite{VCV}.  Of Auger's first 27 events above $\sim 55$ EeV \cite{augerLongAGN}, 19 of the 22 events with $|b|>10^{\circ}$  are within 3.1$^{\circ}$ of a VCV AGN candidate (excluding the Galactic plane where galaxy catalogs are not complete).  

FG09 pointed out that such a correlation with AGNs would be surprising, unless UHECR production occurs in a transient state in which the luminosities of the jet, and the accretion disk powering it, are much higher than found in all but a few of the most powerful steady-state AGNs.  The argument goes as follows.  To correlate within 3.1$^{\circ}$ implies that most UHECRs are protons, because high-Z heavy primaries would be deflected much more than a few degrees in the Galactic magnetic field, and intermediate-Z UHECRs would photodisintegrate enroute from the source.   A proton can only be accelerated to energies such that its Larmor radius is smaller than the size of the accelerator, placing a lower bound on $B \times R$, the magnetic field times source size.   This implies that the (isotropic equivalent) total power in the required magnetic field (Poynting luminosity) is of order
\begin{equation}\label{lumi}
L \sim {1\over 6}c\Gamma ^4B^2R^2 \gtrsim 10^{45} \Gamma ^2\, (E_{20}/Z)^2 \, {\rm erg/s}.
\end{equation}
If the energy in the magnetic field, protons and electrons is in equipartition, and the energy in electrons is emitted through synchrotron cooling in the time it takes the shock to pass through the magnetic cloud, a comparable luminosity is also emitted by electrons with jet-frame Lorentz factors $10^3\lesssim \gamma _e \lesssim  10^8 \, B ^{-\frac{1}{2}}$ placing a limit on the luminosity of the jet.  In order to maintain such a jet, the accretion power should be at least as high, and the bolometric luminosity of the AGN should satisfy Eq. \ref{lumi}.  GRBs easily satisfy this limit, but only the most powerful AGNs satisfy it for $Z=1$.  

Neither GRBs, powerful AGNs, nor arbitrary sources embedded in the local large scale structure (distribution of galaxies as determined from the 2MASS redshift survey) would give rise to a UHECR-AGN correlation on small angular scales at the level reported\cite{zfbMNRAS11}.  Furthermore, FG09 showed from the observed GRB rate and luminosity distribution, that GRBs can only account for the total UHECR flux if the power they emit in UHECRs is much larger than the power emitted in gammas, which would be surprising theoretically.  
FG09 proposed instead that UHECRs are accelerated in jets with a lifetime of order months, produced by TDEs or exceptional accretion disk instabilities.  An association with AGNs would arise if the presence of a thin accretion disk enhanced the probability of a star being captured, or enhanced the probability of the accretion process producing a sufficiently powerful jet. FG09 showed that given the predicted rate of TDEs \cite{wangMerritt04} and a 1\% efficiency, TDFs could easily satisfy both the peak luminoisty requirement, Eq. \ref{lumi}, and the total UHECR energy injection requirement.  

In the following, the SEDs of the two TDFs found in SDSS Stripe 82 archival data by van Velzen et al. (2012) \cite{vf11} (vVF11 below), are used to infer $L_{\rm bol}$ for these TDFs, to learn if TDFs do  indeed satisfy the peak luminoisty requirement, Eq. \ref{lumi}.  In another contribution to this workshop, S. van Velzen and the author %\citet{vfESA12} 
estimate the TDF rate using the vVF11 events.  It is found that TDEs satisfy both requirements for UHECR production: peak luminosity of individual events and adequate total energy injection.   

Before turning to the main task, we note these additional, relevant pieces of information:
\begin{enumerate}
\item Zaw, Farrar and Greene\cite{zfg09} followed up on the VCV galaxies correlating with UHECRs in the first Auger data release.  They found that some of the VCV galaxies were not AGNs, some needed additional observations to classify, and of the ones that are AGNs, most were too weak to satisfy Eq. \ref{lumi}.  
\item Terrano, Zaw and Farrar\cite{tzf12} performed Chandra observations on the correlating VCV galaxies whose status as having an active nucleus remained uncertain, and on one VCV AGN whose bolometric luminosity could not be determined from existing observations\cite{zfg09}; they also classified all but one of the VCV galaxies correlating with UHECRs in the second Auger data release \cite{augerUpdateAGN} and found $L_{\rm bol}$ for all but two cases.  

Combining both Auger published UHECR datasets, there are 57 events with $|b|>10^{\circ}$.  Ref. \cite{tzf12} found that 30-50\% of these events correlate, within the 3.1$^{\circ}$ prescribed by Auger, with genuine AGNs with $z \leq 0.018$.  Of these correlating AGNs, two satisfy Eq. \ref{lumi}, two do not have measured $L_{\rm bol}$, and the rest do not satisfy Eq. \ref{lumi}.  Since about 45\% of the sources of protons above 55 EeV should have $z<0.018$ due to the GZK effect, the observations are consistent with somewhere between roughly half and all of the UHECRs being protons correlating with weak AGNs.
\item  A cluster of 4 UHECRs in the combined AGASA and HiRes data \cite{HRGF} (the ``Ursa Major'' cluster) is consistent with being protons produced in a transient source \cite{gfclus}.  The small angular dispersion of the cluster implies the UHECRs are protons and the absence of a noteworthy source candidate in the field \cite{fbh05} argues that the source is transient. Furthermore, the UHECR energies show the peaked distribution characteristic of a transient source and not the power-law distribution of a continuous source \cite{waxmanME,gfICRC07}. 
\item Auger has measured the distribution of shower maxima, $X_{\rm max}$, in hybrid events up to about 30 EeV.  Taking the predictions of available shower simulations at face value, the observations suggest a shift to heavier composition with increasing energy.  However the simulations fail to describe the observed muon content of the ground showers correctly \cite{jaICRC11}, so it is not yet possible to draw conclusions from the $X_{\rm max}$ distribution; data is also needed at higher energy, because it is the composition of the UHECR dataset which is in question.
\end{enumerate}

\section{Fitting the TDF SEDs}

\begin{figure}[t]
 \centering
 \vspace{-0.75cm}
 \includegraphics[width=0.5\textwidth]{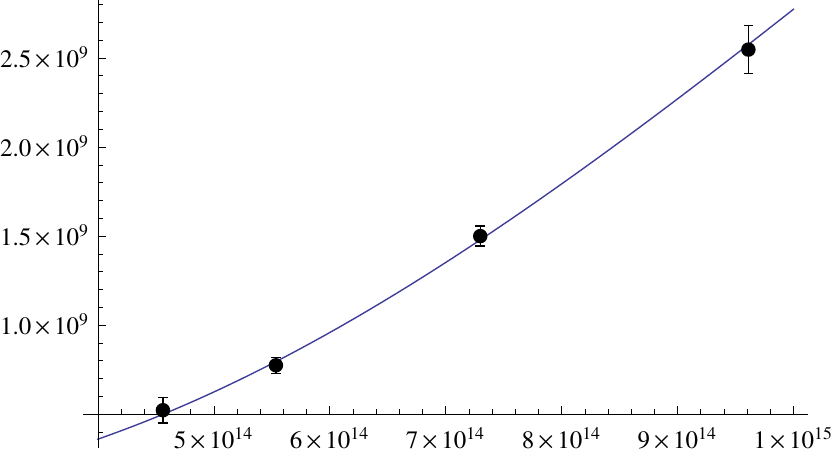}
 %\subfloat[TDE1]{\includegraphics[width=0.5\textwidth]{TDE1datafit.pdf}}
 % \subfloat[TDE2]{\includegraphics[width=0.5\textwidth]{TDE2datafit.pdf}}
 \caption{Thin accretion disk fit to the SED of TDE1; data from vVF11. }
 \label{datafit1}
\end{figure}

\begin{figure}[t]
 \centering
 \vspace{-0.75cm}
 \includegraphics[width=0.5\textwidth]{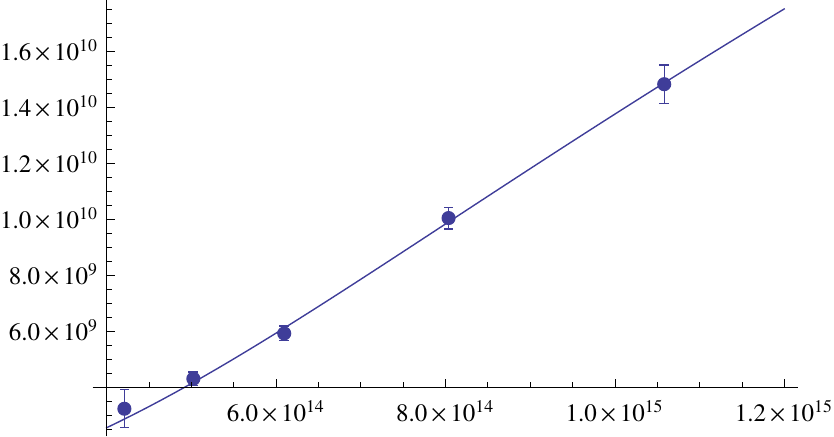}
 %\subfloat[TDE1]{\includegraphics[width=0.5\textwidth]{TDE1datafit.pdf}}
 % \subfloat[TDE2]{\includegraphics[width=0.5\textwidth]{TDE2datafit.pdf}}
 \caption{Thin accretion disk fit to the SED of TDE2; data from vVF11. }
 \label{datafit2}
\end{figure}

The reader is referred to vVF11 for details of the analysis of SDSS Stripe 82 observations which lead to the identification of two tidal disruption events called TDE1 and TDE2, with redshifts $z = 0.136$ and 0.251 respectively.    Their estimated BH masses are (0.6-2) and (2-10) $\times 10^{7 \pm 0.3} M_{\odot}$, where the $\pm 0.3$ in the exponent reflects the scatter in the bulge-BH mass relationship, and the range in prefactors comes from the uncertainty in the bulge masses of the host galaxies.  Figs. \ref{datafit1} and \ref{datafit2} show the SEDs for the two TDEs in the rest frame of their respective BH; the relatively small error bars result from combining observations on different nights by rescaling their mean to the initial observation, there being no significant color evolution over the 3 months of observations.  

For such massive BHs as TDE1,2, the super-Eddington phase is very short, so the observations most likely occured in the accretion-disk dominated phase.  Lodato and collaborators \cite{Lodato09,LR11} and Strubbe and Quataert \cite{SQ09} discuss the light-curves of TDEs and derive the accretion-disk equations in the thin-disk approximation, for an assumed $\dot{M}$.  They find, as one would expect, that the emission is a superposition of black-body annuli, with temperature dropping with radius.  Because the SDSS observations are in the optical, the observed SEDs are dominated by large radii in the accretion disk.  

To restrict the number of free parameters in this initial investigation, the disrupted star was assumed to have the mass and radius of the Sun and default values of accretion efficiency, outflow fraction, etc were adopted from \cite{SQ09}; the ratios of pericenter to tidal disruption radii were set to their most probable values for the given $M_{BH}$'s.  The most important parameters to the fit are the observation time, $M_{BH}$, and outer radius of the accretion disk; these are allowed to vary.  

The quality of both fits is excellent, as can be seen in Figs. \ref{datafit1} and \ref{datafit2}.  The best fit values are $M_{BH} = 1.9 \,\,10^{{7}}\, M_{\odot}  $ and $ 5.6 \,\,10^{7} \, M_{\odot} $, for TDE1,2 respectively, in good agreement with the values inferred from the host galaxies.  The outer radii of the accretion disks are inferred to be 155 and 205 in units of $R_{\rm Sch}$.  The inferred observation epoch is about 6 months for each, consistent with the 9-months off, 3-months on obsevation schedule for Stripe 82 and the fact that the flares were both first seen at the beginning of an observing season.  Integrating over all frequencies gives bolometric luminosities at the time of observation of 5 and $10 \times 10^{47} \, {\rm erg\, s^{-1}}$, respectively.

\section{Conclusions}
We have found that a simple model of thin-disk accretion in a tidal disruption event, based on formulae derived in ref. \cite{SQ09}, gives a good fit to the observed optical spectral energy distributions of both SDSS tidal disruption events.  If this model is valid, only a small fraction of the total luminosity is emitted in optical wavelengths, and $L_{{\rm bol}}$ easily satisfies the criterion for accelerating protons to 100 EeV and above.  

The model does not take into account the temporal evolution of the accretion disk, and numerical simulations will be needed to obtain a more realistic description.  Of particular interest is how plausible it is for the accretion disks to extend to the large radii needed to produce the observed optical emission within this model.   Could such a large radius be a hint of a weak, pre-existing AGN?  (The vVF11 analysis excludes AGNs based on locus in a color-color plot and on variability, but of course a sufficiently tenuous accretion disk could go undetected.

Much more work is needed in modeling the observed SED.  Some of the next steps, in addition to developing more detailed models of the emission, will be to 1) exploit GALEX and Catalina Real-time Transient Survey observations of these two TDEs, which have not been used here,  2) explore more extensively the space of model parameters in the simple thin-disk model, and 3) place upper and lower limits on $L_{\rm bol}$.  Another interesting question is whether a unified model can be developed which describes not only these two SDSS events, but also the recent Swift events -- presumably TDF's viewed from an angle such that jet emission dominates emission from the accretion disk\cite{cenkoSw2058,zaudererSwJ1644}. 

A clearer picture should emerge from UHECR shower observations over the next few years, as to the nuclear composition of UHECRs.  Then, we will know better how stringent the luminosity requirement (\ref{lumi}) on the sources really is.  If a substantial fraction of UHECRs are protons or low-$Z$ nuclei,  bolometric luminosities in excess of $10^{{45}} {\rm erg \, s^{{-1}}} $ will be required and at the same time a rare-source scenario will be ruled out, unless extragalactic magnetic fields produce much larger deflections than the Galactic field does;  at least a portion of UHECRs must in this case be accelerated in transient sources, and Tidal Disruption Flares will be a strong contender.  As more cases like the Ursa Major UHECR cluster with multiple events from a single source are found, whether or not the source is bursting can be determined by the characteristic peaked spectral shape.  

\section{Acknowledgements}
I am grateful to Sjoert van Velzen for a most enjoyable and productive collaboration on the discovery of the two tidal disruption events in SDSS, and other projects, and thank him for providing the SED data used here.  This research has been supported in part by  NSF-PHY-0900631 and NSF-PHY-0970075.  The author is a member of the Pierre Auger Collaboration and thanks her colleagues for their support of and contributions to her research.

\begin{figure}[t]
 \centering
 %\vspace{-0.75cm}
\includegraphics[width=0.5\textwidth]{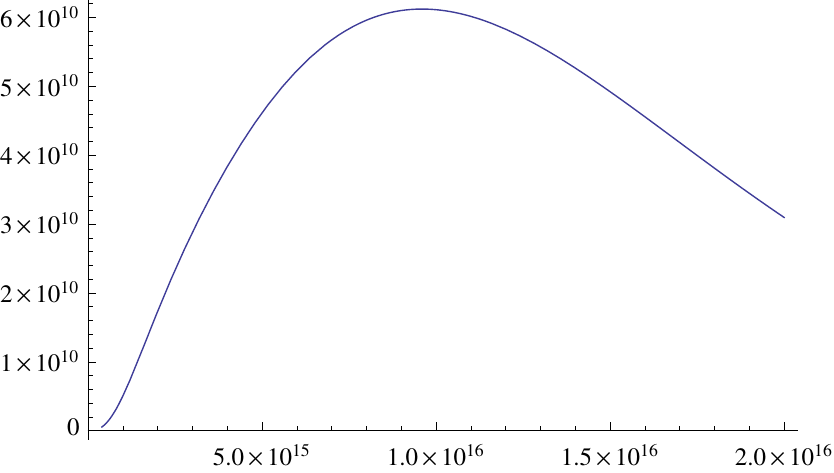}
  % \subfloat[TDE2]{\includegraphics[width=0.5\textwidth]{TDE2nuLnu.pdf}}
 \caption{Full spectral energy distribution of TDE1, using model parameters obtained from the fit shown in Fig. \ref{datafit1} . }
 \label{fullspectrum1}
\end{figure}

\begin{figure}[t]
 \centering
% \vspace{-0.75cm}
\includegraphics[width=0.5\textwidth]{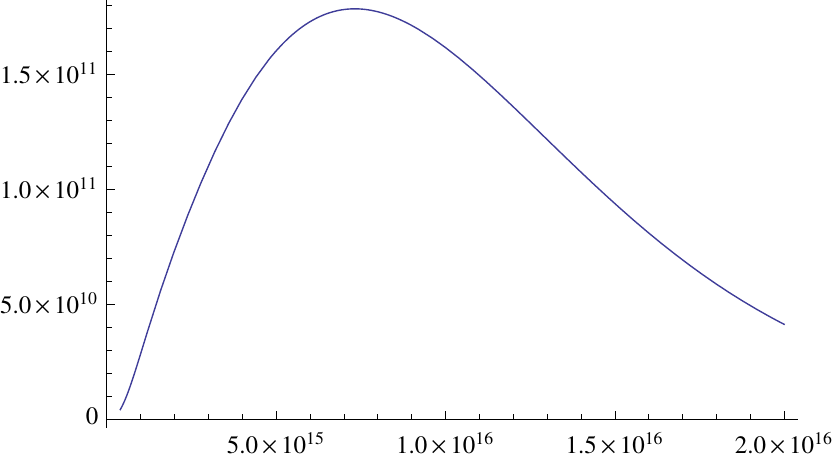}
  % \subfloat[TDE2]{\includegraphics[width=0.5\textwidth]{TDE2nuLnu.pdf}}
 \caption{Full SED of TDE2, using model parameters obtained from the fit shown in Fig. \ref{datafit2} . }
 \label{fullspectrum2}
\end{figure}

%\bibliography{CR,flares}

\end{document}